\documentclass[showpacs,preprintnumbers,amsmath,amssymb]{revtex4}
\usepackage{amsmath,amsfonts,amssymb}
\usepackage{epsfig}
\def\bit{\begin{itemize}}
\def\eit{\end{itemize}}
\def\ben{\begin{enumerate}}
\def\een{\end{enumerate}}
\def\bed{\begin{description}}
\def\eed{\end{description}}

\def\lsim{\raise0.3ex\hbox{$<$\kern-0.75em\raise-1.1ex\hbox{$\sim$}}}
\def\gsim{\raise0.3ex\hbox{$>$\kern-0.75em\raise-1.1ex\hbox{$\sim$}}}

\let\jnfont=\rm
\def\NPB#1,{{\jnfont Nucl.\ Phys.\ B }{\bf #1},}
\def\PLB#1,{{\jnfont Phys.\ Lett.\ B }{\bf #1},}
\def\EPJC#1,{{\jnfont Eur.\ Phys.\ Jour.\ C }{\bf #1},}
\def\PRD#1,{{\jnfont Phys.\ Rev.\ D }{\bf #1},}
\def\PRL#1,{{\jnfont Phys.\ Rev.\ Lett.\ }{\bf #1},}
\def\MPLA#1,{{\jnfont Mod.\ Phys.\ Lett.\ A }{\bf #1},}
\def\JPG#1,{{\jnfont J.\ Phys.\ G}{\bf #1},}
\def\CTP#1,{{\jnfont Commun.\ Theor.\ Phys.\ }{\bf #1},}
\def\JHEP#1,{{\jnfont JHEP \ }{\bf #1},}
\def\NPPS#1,{{\jnfont Nucl.\ Phys.\ Proc.\ Suppl.\ }{\bf #1},}

\def\beq{\begin{equation}}
\def\eeq{\end{equation}}
\def\bea{\begin{eqnarray}}
\def\eea{\end{eqnarray}}
\newcommand{\ba}{\begin{array}}
\newcommand{\ea}{\end{array}}
\def\nn{\nonumber}

\begin{document}
\title{A new regularization of loop integral, no  divergence, no hierarchy problem}

\author{Wenyu Wang$^{1}$, Jian-Feng Wu$^{1}$, Si-Hong Zhou$^{2}$}

\affiliation{
$^1$Institute of Theoretical Physics, College of Applied Science,\\
    Beijing University of Technology, Beijing 100124, China\\
$^2$Institute of High Energy Physics and Theoretical Physics Center for Science Facilities,\\
Chinese Academy of Sciences,Beijing 100049, China}

\begin{abstract}
We find a new regularization scheme which is motivated by the
Bose-Einstein condensation.
The energy of the virtual particle is considered as discrete.
Summing them and regulating the summation by the Riemann $\zeta$ function
can give the result of loop integral.
All the divergences vanish, we can get almost the same results as
Dimensional Regularization.
The prediction beyond Dimensional Regularization
is also shown in the QED. The hierarchy problem of
the radiative correction of scalar mass completely vanish.
\end{abstract}
\maketitle
\section{Introduction}\label{sec1}
It is well known that the higgs suffers from the hierarchy problem.
The key point of the mass hierarchy is the radiative loop of a
scalar with no protection from a symmetry, namely, the one
propagator integration: \beq\int\frac{{\rm d}^d
l}{(2\pi)^d}\frac{i}{l^2-m^2+i\epsilon}\,, \label{int1}\eeq which is
quadratically divergent in four dimension. This implies large fine
tuning between bare value and radiative corrections, making the
higgs in the  Standard model (SM) very unnatural. Here,  we should
suspect on the integral in momentum space more seriously. Always we
use the Fourier transformation between the coordinate and the
momentum space in the study. The basic idea of such transformation
is periodic boundary condition in coordinate or  momentum space and
orthogonality of trigonometrical functions. Summing all the partial
wave gives transformation between periodic functions. Summing
changes to  a continuous integral when the period goes to infinity.
This is all about mathematics. However, it is quite obscure to do
such an integral since in general the energy of a particle in the
quantum world have no reason to be continuous, neither the momenta.

\begin{figure}[htbp]
\scalebox{0.6}{\epsfig{file=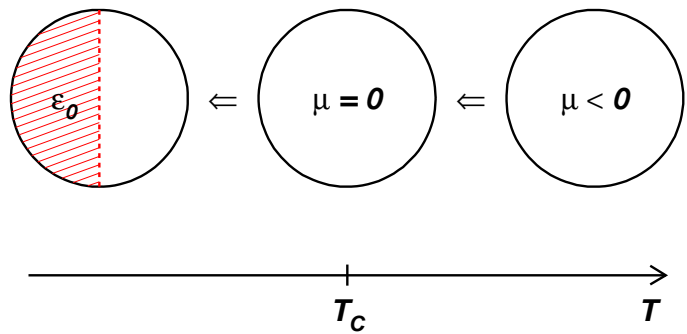}}
\caption{Sketch map  of Bose-Einstein condensation.}
\label{fig1}
\end{figure}
The way from discrete summing to integral is classic in mathematics and physics.
However, such operation may go wrong in some physical systems. One example
is Bose-Einstein condensation (BEC) in statistical physics. The sketch map
of BEC process is shown in FIG. \ref{fig1}: Bose and Einstein worked out the
Bose-Einstein statistics in the analysis of boson system with discrete energy,
then changed it to integral when  they apply the statistics
to a real number density conserved system. When the temperature decreases over
the critical point (the chemical potential $\mu$ approaches zero), finite number of particles
must condense on the ground state which can not be integrated. This is a phase transition
process and there are other physics processes implying
that we must be very careful in the integral operations.
For example, we can do integral to calculate the work of  isothermal processes, but
can not do it for a process composed of infinite infinitesimal adiabatic free expansions,
these two processes can form a superficially same curve in the
pressure and volume plot.

As for the integral Eq. (\ref{int1}) in the momentum space,
we believe the quantum nature should also dominate
the propagator progress such that the integral must be substituted by a summation.
We do not allow propagating particles
have energy lower than its effective mass. This is a physical demand because under
this energy bound, there should be no such a particle. In space-time viewpoint, it means
that for a quantum scattering process triggered,  the collision time should not be too
long because a long time collision implies a classical process happens.
In fact,  a ``point'' or a ``plane wave''  is only a concept in the
mathematical quantum field theory, there are no such things in the real world.
By uncertainty principle, the more precise measurement
on the space-time, the more uncertainty on the momentum-energy.
The increase of the collision energy of the
colliders only increases the accuracy of  the space-time,
but a lower limit of the measurement always exists.
Thus our demand brings a strong constraint on the momentum space and
it implies that there is no way from the summation
to integral. the infinitesimal energy of massive propagating particle does not survive
in a quantum process. There should be a lower bound of smallest energy.

It's very interesting that there are many regularization of divergence summing in mathematics,
especially the Riemann $\zeta(s)$ functions: (Einstein used the finite part in BEC processes.)
$\zeta(s) = \sum_{n=1}^{\infty}  n^{-s}$.
One classical regularization is $\zeta(-1)$:
\beq \zeta(-1) =1+2+3+4+... = -\frac{1}{12}\,,\label{zeta1}\eeq
which details can be found in Ref. \cite{zeta}. Such regularization and BEC process
give us a hint on the integral in the momentum space.
In this paper we find a new regularization scheme of loop integral
by using regularization of Riemann $\zeta$ functions.
The divergences of all the loop integral vanish, and they can be considered as a
condensation on the vacuum partly. Rich physics can be found in our regularization.

The paper is organized in the following: In Sec. \ref{sec2}, we show the detail of
our regularization. In Sec. \ref{sec3} we show the implications of our method.
The conclusion is given in the Sec. \ref{sec4}.
\section{Discrete regularization of loop function}\label{sec2}
The standard method of dimensional regularization (DR) is: Feynman parameterization,
Wick rotation and integral in the Euclidean space. Here we begin at the point after the
Wick rotation. Any loop function in $d$ dimension is
\beq I(d,n,S)=\int\frac{d^{d}l}{(2\pi)^{d}}\frac{1}{(l^{2}+S)^{n}}\,,\label{loop1}\eeq
where $S$ is the effective mass got from  Feynman parameterization. Using DR,
additional $\mu$ parameter should be introduced, the result is:
\beq I_D(d,n,S,\mu)=\mu^{\epsilon}\int\frac{d^{d}l}{(2\pi)^{d}}\frac{1}{(l^{2}+S)^{n}}
=\mu^{\epsilon}\frac{1}{(4\pi)^{d/2}}\frac{\Gamma(n-\frac{d}{2})}
{\Gamma(n)}\biggl(\frac{1}{S}\biggl)^{n-\frac{d}{2}}\,.
\label{loop-dr}\eeq
Note that, $\epsilon$ is a constant, not an input parameter.
However, as argued in the introduction,
 such integral in the energy direction in our scenario should
be replaced by a summation of discrete energy. Thus we consider
the virtual particle like an oscillator, which energy gap is  denoted as $l_0$,
the energy level is  $jl_0$. Integral Eq. (\ref{loop1}) $I(d,n,S)$
 can be changed into another function $I_W(d,n,S,l_0)$ which is
\bea
I_W(d,n,S,l_0)&=&\frac{l_{0}}{\pi}\sum_{j=1}^{\infty}\int\frac{d^{d-1}l}{(2\pi)^{d-1}}
\frac{1}{(l^{2}+j^{2}l_{0}^{2}+S)^{n}}+
\frac{l_{0}}{2\pi}\int\frac{d^{d-1}l}{(2\pi)^{d-1}}
\frac{1}{(l^{2}+S)^{n}}\,.
\eea
Here the gap $l_0$ is a new parameter, the second term is the ground state contribution.
We can see that  this operation can remove one dimension of the divergence, in the following
we will use regularization of Riemann $\zeta$ function to
regulate the divergence. Using Eq. (\ref{loop-dr}) we get:
\bea
I_W(d,n,S,l_0)&=&
\frac{l_{0}}{\pi}\sum_{j=1}^{\infty}\frac{1}{(4\pi)^{d/2-1/2}}
\frac{\Gamma(n-\frac{d}{2}+\frac{1}{2})}{\Gamma(n)}
\biggl(\frac{1}{j^{2}l^{2}_0+S}\biggl)^{n-\frac{d}{2}+\frac{1}{2}}+
\frac{l_{0}}{2\pi}\frac{1}{(4\pi)^{d/2-1/2}}
\frac{\Gamma(n-\frac{d}{2}+\frac{1}{2})}{\Gamma(n)}
\biggl(\frac{1}{S}\biggl)^{n-\frac{d}{2}+\frac{1}{2}}\nn\\
&=& \frac{l_{0}}{\pi}\frac{1}{(4\pi)^{d/2-1/2}}
\frac{\Gamma(n-\frac{d}{2}+\frac{1}{2})}{\Gamma(n)}\biggl[
\sum_{j=1}^{\infty}\left(j^{2}l_{0}^{2}+S\right)^{-(n-\frac{d}{2}+\frac{1}{2})}+
\frac{1}{2}\biggl(\frac{1}{S}\biggl)^{n-\frac{d}{2}+\frac{1}{2}}\biggl]\nn\\
&=&\frac{l_{0}^{-2n+d}}{\pi}\frac{1}{(4\pi)^{d/2-1/2}}
\frac{\Gamma(n-\frac{d}{2}+\frac{1}{2})}{\Gamma(n)}\biggl[
\sum_{j=1}^{\infty}
\left(j^{2}+\frac{S}{l_{0}^{2}}\right)^{-(n-\frac{d}{2}+\frac{1}{2})}+
\frac{1}{2}(S/l_0^2)^{-(n-\frac{d}{2}+\frac{1}{2})}\biggl]\nn\\
&=&\frac{4l_{0}^{-2n+d}}{(4\pi)^{d/2+1/2}}
\frac{\Gamma(n-\frac{d}{2}+\frac{1}{2})}{\Gamma(n)}
\left[E_1^{S/l_0^2}(n-\frac{d}{2}+\frac{1}{2};~1)+
\frac{1}{2}(S/l_0^2)^{-(n-\frac{d}{2}+\frac{1}{2})}\right]\,.\label{wr}
\eea
All the divergences in the $\Gamma$ function now vanish
in case of even number dimension. This is just what we want.  The divergences
are absorbed by the Epstein-Hurwitz function :
\beq
E_1^{c^2}(s;~1)\equiv \sum_{j=1}^{\infty}(j^2+c^2)^{-s}\,,\label{eh}
\eeq
where $c^2=S/l_0^2$ and $s=n-d/2+1/2$.
It can be regulated by Remiann $\zeta$ function in case of $c^{2}\leq1 $,the results
depend on the parameter $s$, which is: \cite{zeta}
\begin{enumerate}
\item  in case of $\frac{1}{2}-s\in N$:
\bea
E_{1}^{c^{2}}(s;1)&=&-\frac{(-1)^{-(s-1/2)}\pi^{1/2}}{2\Gamma(s)\Gamma(\frac{3}{2}-s)}c^{1-2s}
\left[\psi(\frac{1}{2})-\psi(\frac{3}{2}-s)+\ln c^{2}+2\gamma\right]-\frac{1}{2}c^{-2s} \nn\\
&& -\sum_{k=0,k\ne\frac{1}{2}-s}^{\infty}(-1)^{k}\frac{\Gamma(k+s)}{k!\Gamma(s)}\zeta(2k+2s)c^{2k}
\,.\label{numre}
\eea
\item in case of $\frac{1}{2}-s\notin N$ and $-s\notin N$:
\bea
E_{1}^{c^{2}}(s;1)&=& \frac{\pi^{1/2}}{2\Gamma(s)}\Gamma(s-\frac{1}{2})c^{1-2s}-\frac{1}{2}c^{-2s}\nn\\
&& -\sum_{k=0}^{\infty}(-1)^{k}\frac{\Gamma(k+s)}{k!\Gamma(s)}\zeta(2k+2s)c^{2k}\,.\label{anumre}
\eea
\item in case of $-s\in N$:
\bea
E_{1}^{c^{2}}(s;1)&=& -\sum_{k=0}^{-s}(-1)^{k}\frac{\Gamma(k+s)}{k!\Gamma(s)}\zeta(2k+2s)c^{2k}\,.
\eea
\end{enumerate}
where $N$ is the natural number $N=0,1,2,3,\cdots$, $\gamma$ is Euler constant and
$\psi(x)$ is Di-Gamma function. For $d=4$ and $n\le 2$,
we should use the first formula to calculate the loop integral,  for
$n\ge 3$ we should use the second formula. Also we can  use the similar way like dimensional
regularization that we let $d$ approaches the even dimension number, then we should use the
second formula for all the loop integrals. One can check that in case of $n\le 2$, the divergences
of the second formula cancel each other when the dimension number $d$ approaches $4$. In fact,
$E_1^{c^2}(s;~1)$ is a continuous function in the complex plane, thus all the divergences vanish in our
new method of regularization.

Putting the  result of $E_1^{c^2}(s;~1)$ to Eq. (\ref{wr}),
we can get the analytical formula of loop integral
in the new regularization, Note that the second term
of the Eq.(\ref{numre}) and Eq.(\ref{anumre}) cancels
the last term in the bracket of Eq. (\ref{wr}) which is the ground state contribution,
remaining only with two kinds of terms: one is finite term composed by the
product of $\Gamma$ functions, another is a summation of a power series of $S/l_0^2$.
If we set the $l_0^2\gg S$
the first kinds of term will be the leading contribution to the loop integral.
The summation of the power series will be higher order contributions, and it can be evaluated order by order.
Interestingly, the first terms are the exact analytical result of dimensional regularization.  For example,
the result of DR of two point $B_0$ function is: (labeled as $D$)
\bea
B_{0}^{D}&=&\triangle+\mathrm{ln}\frac{\mu^{2}}{m_{1}^{2}}-\int_{0}^{1}\mathrm{d}x\mathrm{ln}S(x)\,,
\eea
in which $\triangle=\frac{2}{\epsilon}-\gamma+\mathrm{ln}4\pi$ is the divergent term
of DR. The result of the new regularization is:  (labeled as $W$)
\bea
B_{0}^{W}&=&2\mathrm{ln}2-2\gamma+\mathrm{ln}\frac{l_{0}^{2}}{m_{1}^{2}}
-\int_{0}^{1}\mathrm{d}x\mathrm{ln}S(x)\nn\\
&& -2\sum_{k=1}^{\infty}(-1)^{k}\frac{\Gamma(k+1/2)}{k!\pi^{1/2}}\zeta(2k+1)
\int_{0}^{1}\mathrm{d}x\left(\frac{m_{1}^{2}}{l_{0}^{2}}S(x)\right)^{k}\,.
\eea
We can see that the divergence regulated from  DR are
replaced by some dimensionless constants. Other
terms of the the first line of both regularization are exactly the same.
Results of other functions are similar, one can check them in the appendix
in which we show the analytical results of A, B and C
function in the two kinds of regularization.

Before using the new regularization, the first question should be
where does the divergence go. As we use the feynman rules
to do the loop calculation, the integration in the momentum  space
is ultraviolet divergent in some functions. In the new regularization,
we use the $\Gamma(x)$ and Riemann $\zeta(s)$ functions
to eliminate the divergences mathematically.
Take quadratically divergent A function as an example,
the divergence are regulated into $\Gamma(-\frac{1}{2})$ and  $\zeta(-1)$ indeed.
These two functions are superficially divergent but are regulated to be finite
by mathematicians. Therefore, we emphasize that the divergence of our new regularization
is only a mathematical problem. However, if using the ordinary dimensional regularization,
the divergences are extracted and canceled by the introducing the counter terms.
Thus they are physical problems, the variables are renormalized by physicists and
the renormalization exposes an understanding of the physical world.
Here we address that there are two levels to understand our regularization:
\begin{enumerate}
\item Level I: This regularization is only a trick by which we can get the almost
               the same results of dimensional regularization in case of $l_0^2\gg S$.
               The divergence are removed and the running of
               renormalization group (shown in the QED) can still remain.
\item Level II: As talked in the introduction, the way from summing to
               integral may go wrong in real physical systems such as in
               the BEC process in which integral should be divergent in case
               of phase transition. What we are doing here
               is kind of an anti-BEC process in which integral
               is divergent but summation became finite.
               We can consider that the divergences are in fact
               condensed in the vacuum. If so,
               we should take a new look at the quantum field theory.
\end{enumerate}
In addition to remove the divergences, another difference between
the DR and the new regularization is the power series term which
become significantly important when $S/l_0^2$ approaches  1.
In the following section, we will show the implication of the
new regularization.
\section{implication of new regularization}\label{sec3}
The quantum field theory has achieved great successes, of which the
most important parts are the radiative corrections and the
renormalizations. At first, physicists do the renormalization just
for dealing with the irritating divergence when doing the radiative
corrections. Finally they found that renormalization exposed a deep
understanding on the quantum theory. In our new regularization,
though there is no divergence in loop integrals, the renormalization
is still necessary, This is because  every physical variable will be
changed by radiative corrections, and we must define a theory at a
renormalization points. Since the leading analytical terms are the
same as the dimensional regularization, the energy gap $l_0$ can be
denoted as the renormalization point of a theory, like the $\mu$
parameter in DR. The renormalization will be exactly the same as the
dimensional regularization when  we set $l_0^2 \gg S$. Such
regularization will be only a trick. Note that though there is no
divergent term, the renormalization group evolution is still the
same in this case. However, here we want to address that what we are
doing is very like an anti-BEC process, thus the energy gap $l_0$ is
kind of the temperature of the vacuum. The verification of our
proposal will be the power series terms of the  loop integral, which
will modify the ordinary loop integral in  case  of the momentum  of
a processes is close to the energy scale at which the theory is
defined. We will see such implication on the quantum
electro-dynamics (QED) and the hierarchy problem of radiative mass
of a scalar.

\subsection{Predications in the QED}
Among all the theories in science, perhaps the QED is the most precisely tested. Loop calculations
are taken into higher and higher order, still no significant deviation is founded. Thus the
new regularization must recover the result of the QED. Also we should use the
precision test of the QED to see that if the new regularization have some new predictions. At one loop
level, the radiative predictions of the QED come from three diagrams:\cite{peskin}
the electron self-energy, the photon self-energy and
the photon-electron vertex. By doing this we can get electron magnetic
moment $a_e$, the Lamb shift, the running of coupling strength $\alpha_{\rm eff }(q^2)$,  and the $\beta$ function
of the QED. As talked above if we use the same terms as in the DR, there is no new prediction
of the new regularization. Here we list and check the modification predicted from
leading term of the power series terms:
\begin{enumerate}
\item {\it Electron magnetic movement $a_e$.}
\bea
a_{e}\equiv\frac{g-2}{2}=\frac{\alpha}{2\pi}
-\frac{\alpha}{2\pi}\frac{\zeta(3)}{6}\frac{m_{e}^{2}}{l_{0}^{2}}\,.
\eea
The first term is the same result of DR and the last term is leading modification from the
new regularization. One  can check that the modification is very small
if $l_0$ is set at electro-weak scale.  More precision measurement of $a_e$ can test
the new regularization. Note that the last term gives a negative contribution to
magnetic movement, thus it can not account for the 3.4$\sigma$ deviation \cite{Hagiwara:2006jt}
of muon magnetic movement by pure QED. However, the complete checking should
include the  electro-weak and hadronic loop contributions
 which is beyond this work.
\item {\it Lamb shift.}
the Lamb shift will not be changed by the new regularization. This is because that
the Uehling potential comes from the imaginary part of photon self energy $\hat\Pi_2(q^2)$
which not appear in the power series terms.
\item {\it Gauge Invariance.}
  Since we divide the one of the four dimension integral into discrete summation,
  the new regulator will violate local U(1) gauge symmetry of the QED,
  thus Ward-Takahashi Identity will no be preserved.
  Among the usual regulators such as Cut-off, Pauli-Villars and DR, only
  DR preserves the gauge symmetry and Lorentz symmetry.
  As talked in the introduction,  continuous gauge symmetry and Lorentz
  symmetry is based on the  ``point'' or  ``plane wave'' mathematical
  quantum field theory, violation of these symmetry in real physics processes
  can be accepted.
\item {\it Running of coupling strength $\alpha_{\rm eff }(q^2)$.}
\bea
\alpha_{\rm eff}(q^{2})=\frac{\alpha}{1-\frac{\alpha}{3\pi}
\mathrm{ln}(-q^{2}/A^{\prime}m^{2})}\,.\label{alprun}
\eea
where $A^{\prime}=\mathrm{exp(\frac{5}{3}+\frac{\zeta(3)}{5})}$. The additional $\zeta$ term
is leading modification from the new regularization. Note that, to get this formula,
we use the condition $l_{0}^{2}\sim-q^{2}\gg m^{2}$
 which means that the energy scale is much higher
than the mass of electron.
\item {\it $\beta$ function of the QED.}
If the energy scale $l_0$ is the temperature of the vacuum, the $\beta$ function
of the coupling is  kind of thermal capacitance of  a theory.
Especially  when the momentum $q^2$ approaches
the temperature $l_0$
($\mathrm{ln}\frac{-q^{2}}{l_{0}^{2}}\sim\mathrm{ln}M^{2}$ e.g. $M^2\to 1$),
 then the $\beta$ function will be exactly the capacitance:
\bea
\beta(\lambda)&=&M\frac{\partial}{\partial M} (\mbox{counter terms})\nn\\
    &=&2\frac{\partial}{\partial \mathrm{ln} M^2 } (\mbox{counter terms})\nn\\
    &\simeq&2\frac{\partial}{\partial M^2 } (\mbox{counter terms})\,.
\eea
This means that the $\beta$ function will change when running
to the high energy scale ($-q^{2}\sim l_{0}^{2}$).
In case of  $-q^2$ approaches  $l_0^2$, for example, $\beta$ function of the QED is:
\bea
\beta(e)=\frac{\mathrm{e}^{3}}{12\pi^{2}}-\frac{1}{3}\zeta(3)\frac{\mathrm{e}^{3}}{16\pi^{2}}\,,\label{beta}
\eea
The first term is the prediction of DR. The second term is the modification of the new regularization.
Using renormalization group equation of the QED
and the Eq. (\ref{beta}), setting $l_0^2=Am^{2}$, we get the running of coupling strength:
\bea
\alpha_{eff}(q^{2})=\frac{\alpha}{1-(\frac{\alpha}{3\pi}-\frac{\zeta(3)\alpha}{12\pi})
\mathrm{ln}(-q^{2}/Am^{2})}\,.
\eea
Similar to the Eq. (\ref{alprun}), the prediction also deviate from result of DR
by $\frac{\zeta(3)\alpha}{12\pi}$ in second term of  denominator.
\end{enumerate}
In all, for the QED,  predictions of the new regularization will be the same as DR at the
low energy. However, when energy increases, the additional power series
term will become significant. When it goes close to the temperature of the vacuum,
the new regularization will give a different story.
\subsection{ Hierarchy problem}
Scalar mass in quantum field theory is a big obstacle of renormalization.
For the radiative correction of a scalar mass is quadratic divergent,
making subtly tuning between bare value and the corrections.
However, in the new regularization, such tuning completely vanishes.
\begin{itemize}
\item {\it$\lambda\phi^4$ theory:} the leading term of mass counter term is
\bea\delta_{m}=\frac{\lambda}{2}\frac{m^{2}}{16\pi^{2}}\left(2\mathrm{ln}2-2\gamma+\mathrm{ln}
\frac{l_{0}^{2}}{m^{2}}+1+\frac{l_{0}^{2}}{3m^{2}}\right)\,.\label{phi4}
 \eea
\item {\it Yukawa theory:} the leading term of mass counter term is
\bea
\delta_{m}=-\frac{Y^{2}}{4\pi^{2}}\left[\frac{l_{0}^{2}}{3}+\int_{0}^{1}dx(m_{f}^{2}-x(1-x)m_{s}^{2})
\left(6\ln2-6\gamma-3\ln \frac{m_{f}^{2}-x(1-x)m_{s}^{2}}{l_0^2}+1\right)\right]+m_{s}^{2}\delta_{Z}\,,
\label{yukawa}
\eea
where $m_f$ is the fermion mass, $m_s$ is the scalar mass and
$\delta_{Z}$ is the counter term of scalar field renormalization which is:
\bea
\delta_{Z}=-\frac{3Y^{2}}{4\pi^{2}}\int_{0}^{1}dxx(1-x)
\left(2\ln2-2\gamma-\frac{2}{3}-\ln \frac{m_{f}^{2}-x(1-x)m_{s}^{2}}{l_0^2}\right)\,.
\eea
\end{itemize}
Though there are quadratic mass term of $l_0^2$ in Eq. (\ref{phi4}) and Eq. (\ref{yukawa}),
$l_0$ (energy gap) is not corresponding to an energy scale of new physics.
Note that the summation of $j$ in Eq. (\ref{wr}) is from 1 to infinity, we do not
set a cut on the integration. What we need is
to set $l_0$ heavier than the effective mass of the integral. In such case,
no tuning is needed and no hierarchy in our new regularization.

We know that the hierarchy problem is perhaps the biggest
theoretical dilemma of SM. It has motivated the most new physics
model beyond SM, such as supersymmetry, extra dimension {\it et.
al.} \cite{np} If we use the new regularization, hierarchy problem
does not seem to be the first motivation of new physics. In our
opinion, the divergence of a radiative correction is non-physical,
the emergence of divergence is because of a wrong mathematical tools
used by physicists. Our regularization gives a clue to the right
tools. Also it exposed a deep understanding of relation between
virtual particle and vacuum. The virtual particles condensed on the
vacuum do not give the radiative corrections. Note that, the
regularization by Riemann $\zeta$ function can also calculate
Casimir effect and the vacuum energy in Ref. \cite{zeta}.
\section{conclusion}\label{sec4}
In this paper, we find a new regularization scheme which is motivated by the BEC process,
the energy of the virtual particle is considered as discrete. Summing them and
regulating the summation by the Riemann $\zeta$ function can
give the result of loop integral.
All the divergences vanish and we can get almost the same results of DR.
The prediction beyond DR is also shown in the QED. The hierarchy problem of
the radiative correction of scalar mass completely vanishes. This give
us very comfortable understanding on the QED and SM in quantum field theory.

As we consider the energy of particle as discrete, one may argue that,
our regularization breaks Lorentz symmetry. However, the leading term
of the new regularization  is exactly the same as DR which preserves the
Lorentz symmetry. The new regularization is very powerful if one treat
it as trick. The Lorentz symmetry violating term may expose a deep understanding
of the vacuum. We want emphasis that what we are doing is the statistics
of vacuum, the implication of the thermal physics of vacuum needs further
study.
\section*{APPENDIX}
Here we list the A, B, C loop function under DR (labeled as $D$)
and the new regularization (label as $W$):
\begin{itemize}
\item A function: $\triangle=\frac{2}{\epsilon}-\gamma+\mathrm{ln}4\pi\,,$
\bea
\frac{i}{16\pi^{2}}A(m^2)&=&\int\frac{\mathrm{d}^dl}{(2\pi)^{d}}
\frac{1}{l^{2}-m^{2}+i\epsilon}\,,
\eea
\bea
A^{D}&=&m^{2}\left[\triangle+\mathrm{ln}\frac{\mu^{2}}{m^{2}}+1\right]\,,\\
A^{W}&=&m^{2}\left[2\mathrm{ln}2-2\gamma+\mathrm{ln}\frac{l_{0}^{2}}{m^{2}}+1\right.\,\nn\\
&& \left. +2\sum_{k=0,k\neq1}^{\infty}(-1)^{k}
\frac{\Gamma(k-\frac{1}{2})}{k!\pi^{1/2}}\zeta(2k-1)\left(\frac{m^{2}}{l_{0}^{2}}\right)^{k-1}\right]\,.
\eea
\item B function: $S(x)=\frac{p^{2}}{m_{1}^{2}}x^{2}-\frac{p^{2}+m_{1}^{2}-m_{2}^{2}}{m_{1}^{2}}x+1\,,$
\bea
\frac{i}{16\pi^{2}}B_{0}(p,m_1^2,m_2^2)&=&{\int\frac{\mathrm{d}^{d}l}{(2\pi)^{d}}
\frac{1}{[l^{2}-m_{1}^{2}+i\epsilon][(l-p)^{2}-m_{2}^{2}+i\epsilon]}}\,,\\
\eea
\bea
B_{0}^{D}&=&\triangle+\mathrm{ln}\frac{\mu^{2}}{m_{1}^{2}}-\int_{0}^{1}\mathrm{d}x\mathrm{ln}S(x)\,,\\
B_{0}^{W}&=&2\mathrm{ln}2-2\gamma+\mathrm{ln}\frac{l_{0}^{2}}{m_{1}^{2}}
-\int_{0}^{1}\mathrm{d}x\mathrm{ln}S(x)\,\nn\\
&& -2\sum_{k=1}^{\infty}(-1)^{k}\frac{\Gamma(k+1/2)}{k!\pi^{1/2}}\zeta(2k+1)
\int_{0}^{1}\mathrm{d}x\left(\frac{m_{1}^{2}}{l_{0}^{2}}S(x)\right)^{k}\,.
\eea
\bea
\frac{i}{16\pi^{2}}B_{\mu}(p,m_1^2,m_2^2)&=&\int\frac{\mathrm{d}^{d}l}{(2\pi)^{d}}
\frac{l_{\mu}}{(l^{2}-m_{1}^{2} +i\epsilon)[(l-p)^{2}-m_{2}^{2}+i\epsilon]}\,,\\
B_{\mu}^{D}&=&\frac{1}{2}p_{\mu}\left[\triangle+\mathrm{ln}\frac{\mu^{2}}{m_{1}^{2}}-2
\int_{0}^{1}\mathrm{d}xx\mathrm{ln}S(x)\right]\,,\\
B_{\mu}^{W}&=&\frac{1}{2}p_{\mu}\left[2\mathrm{ln}2-2\gamma+\mathrm{ln}\frac{l_{0}^{2}}{m_{1}^{2}}-2
\int_{0}^{1}\mathrm{d}xx\mathrm{ln}S(x)\right.\,\nn\\
&& \left.-4\sum_{k=1}^{\infty}(-1)^{k}\frac{\Gamma(k+1/2)}{k!\pi^{1/2}}\zeta(2k+1)
\int_{0}^{1}\mathrm{d}xx\left(\frac{m_{1}^{2}}{l_{0}^{2}}S(x)\right)^{k}\right]\,.
\eea
\bea
\frac{i}{16\pi^{2}}B_{\mu\nu}(p,m_1^2,m_2^2) &=& \int\frac{\mathrm{d}^{d}l}{(2\pi)^{d}}
\frac{l_{\mu}l_{\nu}}{[l^{2}-m_{1}^{2}+i\epsilon] [(l-p)^{2}-m_{2}^{2}+i\epsilon]}\,,
\eea
\bea
B_{\mu\nu}^{D} &=& \frac{1}{3}\left\{p_{\mu}p_{\nu}\left[\Delta+\mathrm{ln}
\frac{\mu^{2}}{m_{1}^{2}}-3\int_{0}^{1}\mathrm{d}xx^{2}\mathrm{ln}S(x)\right]\right.\,\nn\\
&&
+\frac{g_{\mu\nu}}{d}\left[(3m_{1}^{2}+3m_{2}^{2}-p^{2})(\Delta+\mathrm{ln}
\frac{\mu^{2}}{m_{1}^{2}}+1)\right.\,\nn\\
&&
\left.\left.-\frac{3m_{1}^{2}+3m_{2}^{2}-p^{2}}{2}-6m_{1}^{2}\int_{0}^{1}\mathrm{d}xS(x)
\mathrm{ln}S(x)\right]\right\}\,,
\eea
\bea
B_{\mu\nu}^{W}& =&\frac{1}{3}\left\{p_{\mu}p_{\nu}\left[2\mathrm{ln}2-2\gamma+\mathrm{ln}
\frac{l_{0}^{2}}{m_{1}^{2}}
-3\int_{0}^{1}\mathrm{d}xx^{2}\mathrm{ln}S(x)\right]\right.\,\nn\\
&&
 +\frac{g_{\mu\nu}}{d}\left[(3m_{1}^{2}+3m_{2}^{2}-p^{2})(2\mathrm{ln}2-2\gamma
+\mathrm{ln}\frac{l_{0}^{2}}{m_{1}^{2}}+1)\right.\,\nn\\
&&
\left.-\frac{3m_{1}^{2}+3m_{2}^{2}-p^{2}}{2}-6m_{1}^{2}
\int_{0}^{1}\mathrm{d}xS(x)\mathrm{ln}S(x)\right]\,\nn\\
&&
-6p_{\mu}p_{\nu}\left[\sum_{k=1}^{\infty}(-1)^{k}\frac{\Gamma(k+1/2)}{k!\pi^{1/2}}\zeta(2k+1)
\int_{0}^{1}\mathrm{d}xx^{2}\left(\frac{m_{1}^{2}}{l_{0}^{2}}S(x)\right)^{k}\right]\,\nn\\
&&
-6\frac{g_{\mu\nu}}{d}\left[-l_{0}^{2}\sum_{k=0,k\neq1}^{\infty}(-1)^{k}\frac{\Gamma(k-1/2)}
{k!\pi^{1/2}}\zeta(2k-1)\int_{0}^{1}\mathrm{d}x\left(\frac{m_{1}^{2}}{l_{0}^{2}}S(x)\right)^{k}\right.
\,\nn\\
&&
+\left.\left.\sum_{k=1}^{\infty}(-1)^{k}\frac{\Gamma(k+1/2)}{k!\pi^{1/2}}\zeta(2k+1)\int_{0}^{1}\mathrm{d}x
m_{1}^{2}S(x)\left(\frac{m_{1}^{2}}{l_{0}^{2}}S(x)\right)^{k}\right]\right\}\,.
\eea
\item C function: $S(x,y)=(x^{2}p^{2}+y^{2}p^{\prime2}+2xypp^{\prime})-x(p^{2}+m_{1}^{2}-m_{2}^{2})
-y(p^{\prime2}+m_{1}^{2}-m_{3}^{2})+m_{1}^{2}\,,$
\bea
\frac{i}{16\pi^{2}}C_{0}(p,p',m_1^2,m_2^2,m_3^2)
&=&\int\frac{\mathrm{d}^{d}l}{(2\pi)^{d}}\frac{1}{(l^{2}-m_{1}^{2}+i\epsilon)
[(l-p)^{2}-m_{2}^{2}+i\epsilon][(l-p^{\prime})^{2}-m_{3}^{2}+i\epsilon]}\,,
\eea
\bea
C_{0}^{D};C_{\mu}^{D}&=&-\int_{0}^{1}\mathrm{d}x\int_{0}^{1-x}
\mathrm{d}y(1;xp_{\mu}+yp_{\mu}^{\prime})\frac{1}{S(x,y)}\,,
\eea
\bea
C_{0}^{W};C_{\mu}^{W}&=&-\int_{0}^{1}\mathrm{d}x\int_{0}^{1-x}
\mathrm{d}y(1;xp_{\mu}+yp_{\mu}^{\prime}) \left[\frac{1}{S(x,y)}\right.\,\nn\\
&&
-\left.\sum_{k=0}^{\infty}(-1)^{k}\frac{2\Gamma(k+\frac{3}{2})}{k!\pi^{1/2}}
\zeta(2k+3)l_{0}^{-2}\left(\frac{S(x,y)}{l_{0}^{2}}\right)^{k}\right]\,.
\eea
\bea
\frac{i}{16\pi^{2}}C_{\mu\nu}(p,p',m_1^2,m_2^2,m_3^2)&=&\int\frac{\mathrm{d}^{d}l}{(2\pi)^{d}}
\frac{l_{\mu}l_{\nu}}{(l^{2}-m_{1}^{2}-i\epsilon)
[(l-p)^{2}-m_{2}^{2}+i\epsilon][(l-p^{\prime})^{2}-m_{3}^{2}-i\epsilon]}\,,
\eea
\bea
C_{\mu\nu}^{D}&=&\frac{g_{\mu\nu}}{4}\triangle-\int_{0}^{1}\mathrm{d}x\int_{0}^{1-x}\mathrm{d}y
\left[\frac{x^{2}p_{\mu}p_{\nu}+xy(p_{\mu}p_{\nu}^{\prime}
+p_{\mu}^{\prime}p_{\nu})+y^{2}p_{\mu}^{\prime}p_{\nu}^{\prime}}{S(x,y)}
+\frac{g_{\mu\nu}}{2}\mathrm{ln}\frac{S(x,y)}{\mu^{2}}\right]\,,
\eea
\bea
C_{\mu\nu}^{W}&=&\frac{g_{\mu\nu}}{4}(2\mathrm{ln}2-2\gamma-\frac{1}{2})\,\nn\\
&& -\int_{0}^{1}\mathrm{d}x\int_{0}^{1-x}\mathrm{d}y\left[\frac{x^{2}p_{\mu}p_{\nu}
+xy(p_{\mu}p_{\nu}^{\prime}+p_{\mu}^{\prime}p_{\nu})+y^{2}p_{\mu}^{\prime}p_{\nu}^{\prime}}{S(x,y)}+
\frac{g_{\mu\nu}}{2}\mathrm{ln}\frac{S(x,y)}{l_{0}^{2}}\right]\,\nn\\
&& +\frac{g_{\mu\nu}}{2}\left[-2\sum_{k=1}^{\infty}(-1)^{k}\frac{\Gamma(k+1/2)}{k!\pi^{1/2}}
\zeta(2k+1)\int_{0}^{1}\mathrm{d}x\right.
\int_{0}^{1-x}\mathrm{d}y\left(\frac{S(x,y)}{l_{0}^{2}}\right)^{k}\, \nn\\
&&
+\left.\frac{1}{2}\sum_{k=0}^{\infty}(-1)^{k}\frac{2\Gamma(k+\frac{3}{2})}{k!\pi^{1/2}}\zeta(2k+3)\int_{0}^{1}\mathrm{d}x
\int_{0}^{1-x}\mathrm{d}y\left(\frac{S(x,y)}{l_{0}^{2}}\right)^{k+1}\right]\,\nn\\
&&
+\sum_{k=0}^{\infty}(-1)^{k}\frac{2\Gamma(k+\frac{3}{2})}{k!\pi^{1/2}}\zeta(2k+3)
\int_{0}^{1}\mathrm{d}x\int_{0}^{1-x}\mathrm{d}y\,\nn\\
&& \times\left[x^{2}p_{\mu}p_{\nu}+xy(p_{\mu}p_{\nu}^{\prime}+p_{\mu}^{\prime}p_{\nu})
+y^{2}p_{\mu}^{\prime}p_{\nu}^{\prime}\right]l_{0}^{-2}\left(\frac{S(x,y)}{l_{0}^{2}}\right)^{k}\,.
\eea
\bea
\frac{i}{16\pi^{2}}C_{\mu\nu\sigma}(p,p',m_1^2,m_2^2,m_3^2)&=&\int\frac{\mathrm{d}^{d}l}{(2\pi)^{d}}
\frac{l_{\mu}l_{\nu}l_{\sigma}}{(l^{2}-m_{1}^{2}+
i\epsilon)[(l-p)^{2}-m_{2}^{2}+i\epsilon][(l-p^{\prime})^{2}-m_{3}^{2}+i\epsilon]}\,,
\eea
\bea
C_{\mu\nu\sigma}^{D}&=&\frac{\Delta}{12}\left[g_{\mu\nu}(p+p^{\prime})_{\sigma}+g_{\nu\sigma}
(p+p^{\prime})_{\mu}+g_{\sigma\mu}(p+p^{\prime})_{\nu}\right]\,\nn\\
&&
-\int_{0}^{1}\mathrm{d}x\int_{0}^{1-x}\mathrm{d}y\biggl\{\left[x^{3}p_{\mu}p_{\nu}
p_{\sigma}+x^{2}y(p_{\mu}p_{\nu}p_{\sigma}^{\prime}+
p_{\mu}p_{\nu}^{\prime}p_{\sigma}+p_{\mu}^{\prime}p_{\nu}p_{\sigma})\right.\,\nn\\
&&
+\left.xy^{2}(p_{\mu}^{\prime}p_{\nu}^{\prime}p_{\sigma}+p_{\mu}^{\prime}p_{\nu}p_{\sigma}^{\prime}
+p_{\mu}p_{\nu}^{\prime}p_{\sigma}^{\prime})
+y^{3}p_{\mu}^{\prime}p_{\nu}^{\prime}p_{\sigma}^{\prime}\right]/S(x,y)\,\nn\\
&&
+\frac{1}{2}\left[g_{\mu\nu}(xp+yp^{\prime})_{\sigma}+g_{\nu\sigma}(xp+yp^{\prime})_{\mu}+g_{\sigma\mu}
(xp+yp^{\prime})_{\nu}\right]\mathrm{ln}\frac{S(x,y)}{l_{0}^{2}}\biggl\}\,,
\eea
\bea
C_{\mu\nu\sigma}^{W}&=&\frac{2\mathrm{ln}2-2\gamma-\frac{1}{2}}{12}[g_{\mu\nu}(p+p^{\prime})_{\sigma}
+g_{\nu\sigma}(p+p^{\prime})_{\mu}+g_{\sigma\mu}(p+p^{\prime})_{\nu}]\,\nn\\
&&
-\int_{0}^{1}\mathrm{d}x\int_{0}^{1-x}\mathrm{d}y\biggl\{\left[x^{3}p_{\mu}p_{\nu}p_{\sigma}+
x^{2}y(p_{\mu}p_{\nu}p_{\sigma}^{\prime}+p_{\mu}p_{\nu}^{\prime}p_{\sigma}
+p_{\mu}^{\prime}p_{\nu}p_{\sigma})\right.\,\nn\\
&&
+\left.xy^{2}(p_{\mu}^{\prime}p_{\nu}^{\prime}p_{\sigma}+p_{\mu}^{\prime}p_{\nu}p_{\sigma}^{\prime}+
p_{\mu}p_{\nu}^{\prime}p_{\sigma}^{\prime})+y^{3}p_{\mu}^{\prime}p_{\nu}^{\prime}
p_{\sigma}^{\prime}\right]/S(x,y)\,\nn\\
&&
+\frac{1}{2}\left[g_{\mu\nu}(xp+yp^{\prime})_{\sigma}+g_{\nu\sigma}(xp+yp^{\prime})_{\mu}+g_{\sigma\mu}
(xp+yp^{\prime})_{\nu}\right]\mathrm{ln}\frac{S(x,y)}{l_{0}^{2}}\biggl\}\,\nn\\
&&
-\int_{0}^{1}\mathrm{d}x\int_{0}^{1-x}\mathrm{d}y\left\{\frac{1}{2}\left[g_{\mu\nu}
(xp+yp^{\prime})_{\sigma}+g_{\nu\sigma}(xp+yp^{\prime})_{\mu}+g_{\sigma\mu}
(xp+yp^{\prime})_{\nu}\right]\right.
\,\nn\\
&&
 \times \left[2\sum_{k=1}^{\infty}(-1)^{k}\frac{\Gamma(k+1/2)}{k!\pi^{1/2}}\zeta(2k+1)
\left(\frac{S(x,y)}{l_{0}^{2}}\right)^{k}
 -\frac{1}{2}\sum_{k=0}^{\infty}(-1)^{k}\frac{2\Gamma(k+\frac{3}{2})}{k!\pi^{1/2}}\zeta(2k+3)
 \left(\frac{S(x,y)}{l_{0}^{2}}\right)^{k+1}\right]\,\nn\\
 &&
 +\left[x^{3}p_{\mu}p_{\nu}p_{\sigma}+x^{2}y(p_{\mu}p_{\nu}p_{\sigma}^{\prime}+
  p_{\mu}p_{\nu}^{\prime}p_{\sigma}+p_{\mu}^{\prime}p_{\nu}p_{\sigma})
 +xy^{2}(p_{\mu}^{\prime}p_{\nu}^{\prime}p_{\sigma}+p_{\mu}^{\prime}p_{\nu}p_{\sigma}^{\prime}
+p_{\mu}p_{\nu}^{\prime}p_{\sigma}^{\prime})
 +y^{3}p_{\mu}^{\prime}p_{\nu}^{\prime}p_{\sigma}^{\prime}\right]\,\nn\\
 &&  \times\sum_{k=0}^{\infty}(-1)^{k}\frac{2\Gamma(k+\frac{3}{2})}
 {k!\pi^{1/2}}\zeta(2k+3)\left.l_{0}^{-2}\left(\frac{S(x,y)}{l_{0}^{2}}\right)^{k}\right\}\,.
\eea
\end{itemize}

\end{document}